\documentstyle[prl,aps,twocolumn]{revtex}

\begin{document}

\title{Silver Filled Carbon Nanotubes used as Spectroscopic Enhancers}

\author{F.J. Garc\'{\i}a-Vidal,$^1$ J.M. Pitarke,$^2$ and J.B. Pendry$^3$}

\address{
$^1$ Departamento de F\'{\i}sica 
Te\'orica de la Materia Condensada. Facultad de Ciencias, \\
Universidad 
Aut\'onoma de Madrid. Madrid 28049 (Spain). \\
$^2$ Materia Kondentsatuaren Fisika
Saila. Zientzi Fakultatea. \\
Euskal Herriko Unibersitatea. 644 Posta kutxatila. 48080 Bilbo
(Spain).\\
$^3$ Condensed Matter Theory Group, 
The Blackett Laboratory, Imperial College, 
London SW7 2BZ (UK).}

\maketitle

\begin{abstract}
We analyse from a theoretical point of view the optical properties of arrays
of carbon nanotubes filled with silver. Dependence of these properties on
the different parameters involved is studied using a Transfer Matrix
formalism able to work with tensor-like dielectric functions and including
the full electromagnetic coupling between the nanotubes. We find these
structures exhibit very strong linear optical response and hence could be
used as spectroscopic enhancers or chemical sensors in the visible range.
Very localised surface plasmons, created by the electromagnetic interaction
between the capped silver cylinders, are responsible of this enhancing
ability. Enhancements of up to $10^6$ in the Raman signal of molecules
absorbed on these arrays could be obtained. \newline
\end{abstract}


The so called coinage metals (Au, Ag and Cu) exhibit very strong linear and
non-linear optical response when they are structured on the nanometre scale.
Their ability to enhance optical fields has been widely used in the last
twenty years to study spectra of several molecules, taking advantage of the
Surface Enhanced Raman Scattering (SERS) effect \cite{sers}. This effect
appears on suitably rough metal surfaces of Au, Ag or Cu and is connected
with excitation of very localised surface plasmons present in the vicinity
of highly coupled metal features \cite{yo}. However, controlled use of
coinage metals as spectroscopic enhancers has proven to be very difficult,
due to the absence of well-defined, stable macroscopic materials made of
them, and various approaches to preparation of
ordered arrays of metal nanoparticles have been reported only very recently 
\cite{dots}.
Basically, these techniques consist of capping the nanoparticles with organic
molecules of different sizes in order to control the interparticle
separation.

Carbon nanotube technology is another promising new area in materials
science \cite{carbon}. Multishell nanotubes made of a number of concentric
cylinders of planar graphite with diameters ranging from 2 to 25 nm and
lengths of up to several microns have now become available. One of the most
interesting applications of these nanometre size tubes is linked to their
hollow character: carbon nanotubes could be used as nanoscale moulds. Filled
nanotubes can be synthesised by capillary action \cite{capi} or by using
composite anodes in the arc discharge \cite{anodes}, resulting in formation
of encapsulated compounds or elongated nanostructures of different materials
(carbides, oxides, organic solvents or even pure metals). Very recently it
has been possible to fill carbon nanotubes with a coinage metal like silver,
using capillary forces \cite{ugarte}. On the other hand, bulk alignment of
nanotubes has been also reported using different techniques \cite
{films,deheer}. In these ordered arrays the carbon nanotubes form very
close-packed structures.

The aim of this letter is to propose a scenario that combines both areas of
materials science. We investigate the optical response of arrays of
Ag-filled carbon nanotubes in order to see if the capacity of silver
nanostructures to enhance optical fields is retained when they are
encapsulated inside carbon nanotubes. If so, these well-defined structures
could have important optical applications as, for example, spectroscopic
enhancers or chemical sensors in the visible range.

Our model for an array of Ag-filled carbon nanotubes is displayed in figure
1. The tubes are infinitely long in the $z$ direction and form a periodic
structure in the $x$ direction. In accordance with experimental evidence
\cite{ugarte,deheer}, the outer diameter of the nanotubes is chosen to be $%
d_{out}=10$ nm and the distance between them, $d$, $10.3$ nm
(close-packed structure). It has been reported that only tubes with inner
diameters of $4$ nm or more can be filled with silver using capillary forces,
and the distribution of inner diameters is broad varying from $4$ to $9$ nm
\cite{ugarte}. Bearing this in mind, the diameter of the encapsulated silver
tubules, $d_{in}$, will be varied in our calculations between $3$ and $9.5$
nm, in order to study dependence of the optical response of these arrays on 
the diameter of the inner core. Since the outer diameter of the tubes and
the distance between them are fixed, varying $d_{in}$ changes the separation
between metal nanoparticles.

We consider an electromagnetic p-polarised plane
wave of frequency $\omega$ and momentum ${\bf {k}}$ normally incident on 
the structure, and we analyse its 
interaction with an array of
carbon nanotubes. For this polarisation, the incident {\bf E} field is perpendicular 
to the axis of the tubes and surface plasmons of 
the nanotubes can be excited. In particular, we calculate the optical 
reflectance,
$R({\bf {k}},\omega )$, of these structures and the possible enhancement of
the electric field at various positions on the surface. For both quantities
we first need to compute the reflection matrices, $\hat{R}({\bf {k}},{\bf
{k}%
^{\prime }};\omega )$, defining scattering of the incoming wave into outgoing
waves of momenta ${\bf {k}^{\prime }}$. When dealing with materials whose
dielectric response disperses strongly with frequency, such as graphite and
silver, on-shell methods like the one developed in our group \cite{John} are
ideally suited to calculate these matrices. First we fix $\omega $ and hence
$\epsilon (\omega )$ can be specified. By approximating the continuous
fields by their values at a series of discrete points, we can construct the
EM transfer matrix of our system. This transfer matrix, relating EM fields
at one side of the structure to those on the other side in the $y$
direction, allows us to calculate transmission and reflection coefficients
for an incoming plane wave \cite{John}. Once we have these matrices, the
reflectance of the surface can be easily calculated:

\begin{equation}
R({\bf {k}},\omega )=\sum_{{\bf k}^{\prime }}|\hat{R}({\bf {k}},{\bf {k}%
^{\prime }};\omega )|^2,
\end{equation}
\noindent
where the sum runs over only propagating outgoing waves.

In order to calculate the electric field surrounding the carbon nanotubes,
we first construct the total ${\bf E}$ field outside the surface as a sum of
the incident and reflected waves,

\begin{eqnarray}
{\bf E}_{{\rm total}}({\bf {r},}\omega )=e^{i{\bf k\cdot r}} &&{\bf {E}}%
_{inc}({\bf k},\omega ) \\
&&+\sum_{{\bf k}^{\prime }}e^{i{\bf {k}^{\prime }}\cdot {r}}\hat{R}({\bf
k},%
{\bf k}^{\prime };\omega ){\bf {E}(k^{\prime }},\omega ),  \nonumber
\end{eqnarray}
\noindent
where ${\bf {E}}_{inc}({\bf k},\omega)$ is the electric field associated
with an incoming plane wave and ${\bf {E}({k}^{\prime },\omega )}$ is the
electric field associated with the outgoing plane wave ${\bf
{k^{\prime }}}$%
. In this case the sum runs over both propagating and evanescent waves. Then
we integrate the EM fields in real space through the system to obtain a
detailed picture of the electric field.

As we said above, multishell nanotubes are made of several concentric
cylinders of planar graphite. Graphite is highly anisotropic and it is
necessary to distinguish two different components, $\epsilon _{\parallel
}(\omega )$ and $\epsilon _{\perp }(\omega )$, in its dielectric function
for the directions parallel and normal to the axis, respectively. When
transferring these dielectric functions to cylindrical geometry, for every
point $(x,y)$ inside the nanotube and outside the inner core we can write
down a local dielectric tensor \cite{Lucas,nano1},

\begin{equation}
\hat{\varepsilon}(x,
y,\omega)=\left( 
\begin{array}{ccc}
\frac{x^2}{r^2}\varepsilon_{\parallel}+ \frac{y^2}{r^2}\varepsilon_{\perp} & 
\frac{xy}{r^2}(\varepsilon_{\parallel}-\varepsilon_{\perp}) & 0 \\ 
\frac{xy}{r^2}(\varepsilon_{\parallel}-\varepsilon_{\perp}) & \frac{y^2}{r^2}%
\varepsilon_{\parallel}+ \frac{x^2}{r^2}\varepsilon_{\perp} & 0 \\ 
0 & 0 & \varepsilon_{\perp}
\end{array}
\right),
\end{equation}
\noindent 
where $r=\sqrt{x^2+y^2}$. For a point inside the inner core the local
dielectric tensor is diagonal and equal to the dielectric function of
silver, $\epsilon_{Ag}(\omega)$. For the calculations we show in this paper
we have used the dielectric functions of graphite ($\epsilon_{\parallel}(
\omega)$ and $\epsilon_{\perp}(\omega)$) and silver ($\epsilon_{Ag}(\omega)$%
), as tabulated in Ref.\onlinecite{Palik}. 

Very recently, an extension of the Transfer Matrix formalism described above
has been developed \cite{Andrew}, which is able to solve Maxwell equations
with tensor-like dielectric functions like the one present in Eq. (3). This
method has been applied with success \cite{nano1} to explain the reported
optical data of aligned carbon nanotube films \cite{deheer}. Within this
on-shell formalism it is possible to compute the reflection matrices needed
to calculate the optical reflectance and the electric fields in real space.

The linear optical response
of silver and other coinage metals is dominated
by the surface plasmon resonance, $\omega _{sp}$. For a nanometric particle,
$\omega _{sp}$ depends on the plasma frequency of the metal, the dielectric
surroundings and the particular geometry of the nanostructure. For an {\it %
isolated} metal cylinder in vacuum, the location of $\omega _{sp}$ is
determined by the equation $\epsilon (\omega )+1=0$ and the surface plasmon
has a dipolar character. For the case of a silver cylinder $\omega _{sp}$ $%
=3.6$ eV. When nanoparticles are brought into close contact, electromagnetic
coupling between them provokes a shift of $\omega _{sp}$ to lower energies
and the dipolar surface plasmon converts into a localised one, trapped in
the region between the nanostructures \cite{yo}. An incident plane wave of
appropriate frequency can excite this surface plasmon, creating a huge {\bf
E%
} field at this location and leading to large reductions in optical reflectance
at this frequency.

In Fig. 2 we show the optical reflectance
for frequencies ranging from $1$ to $4$ eV, calculated for an array of
close-packed carbon nanotubes filled with silver (see Fig.1). The diameter
of the inner core, $d_{in}$, is varied between $3$ and $9.5$ nm. As shown in
the figure, for $d_{in}\ge 5$ nm there is a dip in the frequency dependence
of the optical reflectance. This is a clear signal that for these values of
$%
d_{in}$, the incident photon is exciting surface plasmons associated with
the silver cylinders filling the nanotubes. Plasmons of carbon are located
at higher energies, $5.2$ and $6.2$ eV, and cannot be excited at these
frequencies \cite{nano1}. Moreover, the shift to lower energies of $\omega
_{sp}$ as $d_{in}$ is increased indicates that encapsulated silver
nanostructures are electromagnetically coupled, their carbon shells serving
as the medium for this interaction. We note that the silver surface plasmon 
is absent when 
$d_{in}$ is less than $4$ nm, just the
minimum value of $d_{in}$ at which a carbon nanotube can be filled with
silver using capillary forces \cite{ugarte}. Another interesting conclusion
that can be drawn from looking at this figure is that the shift to lower
energies of $\omega _{sp}$ is accompanied by a narrowing of the plasmon
linewidth. A narrow linewidth indicates that radiative coupling to the
vacuum of the plasmon is weak, and therefore a huge enhancement of the
optical fields is expected for the largest values of $d_{in}$.

It can be shown that for small values of $d_{in}$ ($5$ or $6$ nm) the dip in
the frequency dependence of the optical reflectance is associated with
excitation of a {\bf dipolar} surface plasmon, similar to the one present in
{\it isolated} silver cylinders. However, as the distance between the silver
cylinders is reduced, the surface plasmon excited becomes more localised. In
Fig. 3, we show a detailed picture of the {\bf E} field generated by the
incident photon for $d_{in}=8$ nm. We have evaluated the fields at
$w=2.5$ eV, the frequency of minimum reflectance for this case (see Fig. 2).
Clearly the incident radiation is exciting a very localised surface plasmon
and the intensity of the {\bf E} field is strongly enhanced with respect to
the incident one.
The electric field within the carbon shell is also large because at this
range of frequencies carbon behaves effectively as a lossy dielectric. It is
also worth mentioning that the induced charge (just the divergence of the
{\bf E} field shown in Fig. 3) is located not only on the surface of the
silver cylinders but also within the carbon shell. This is due to the 
tensor-like character of the optical response of graphite that allows to 
fulfil the condition $\nabla \cdot (\hat{\varepsilon} {\bf E})=0$ with 
$\nabla \cdot {\bf E}$ different form zero not only at the interfaces but also 
inside the nanotubes. This {\it bulk} induced
charge reinforces the intensity of the {\bf E} field created by excitation
of the silver surface plasmon.

These results demonstrate that the ability to enhance optical fields,
inherent to silver nanostructures, is still present when they are
encapsulated inside carbon nanotubes. However, as the surface plasmon
strength depends on the diameter of the silver inner core, this capability
is expected to be very sensitive to this diameter. In order to evaluate this
enhancing property quantitatively, and investigate the possibility of using
these surfaces as SERS active substrates, we have calculated the local
enhancement of the Raman signal of a molecule absorbed in the region between
the nanotubes. To a first approximation, this Raman signal depends on the
fourth power of the total electric field at the molecule position (${\bf
E}(%
{\bf r}_m,\omega )$), and hence its enhancement due to the presence of the
surface is simply:

\begin{equation}
\rho \left( {\bf r}_{m}, \omega \right) = \left| {\frac{{{\bf E} ({\bf r}%
_{m},\omega)} }{{\ E_{{\rm inc}}(\omega)}}} \right|^4 \, ,
\end{equation}
\noindent
where ${\bf r}_m$ represents the location of the molecule, and $E_{{\rm 
inc}}(\omega
)$ is the electric field associated with the incident plane wave.

In Fig. 4 this enhancement is shown, as a function of the
incoming photon energy and for different values of $d_{in}$, ranging from $5$
to $9.5$ nm. 
As expected from dependence of the excited surface plasmon
linewidth on $d_{in}$, enhancement grows rapidly from $%
10^4$ for $d_{in}=5$ nm to $10^6$ for $d_{in}=9.5$ nm. As $d_{in}$ is
increased, the induced charge is becoming more localised and, therefore, the
{\bf E} field in the region between the nanotubes is larger. This behaviour
had previously been found when studying the optical response of silver
nanoparticles deposited on a silver surface \cite{yo}. For comparison, we 
show in the same figure the expected enhancement of the Raman signal for  
an array of bare silver cylinders with diameters of $9.5$ nm, their axis being 
separated by a distance of $10.3$ 
nm. 
Due to the lossy characteristics of carbon at this  
range of frequencies, encapsulated silver nanostructures have a weaker enhancing 
capacity with respect to the bare ones. On the other hand, 
in the case of encapsulated tubes the  
dependence of the enhancement on the energy of the incoming photon
has a less pronounced resonant behaviour, and its value is large 
and practically constant   
at all frequencies within the visible range. 
This distinctive property of Ag-filled carbon 
nanotubes might be expected to be useful in practical applications. 

In conclusion, we have investigated the optical response of carbon
nanotubes filled with silver. We have shown how a silver nanostructure's
ability to enhance optical fields is retained when it is encapsulated inside
these tubes. For the largest values of inner core radius,  
the enhancement of the {\bf E} field intensity can be as large as $10^3$.
The experimental viability of building up ordered 
structures made of these silver filled carbon tubes could open the possibility of 
using these 
systems as spectroscopic enhancers in the visible range. 
The optical properties of these structures could be tuned by 
varying 
the diameter of the inner core, carbon shells playing the role of moulds 
that avoid aggregation between the metal nanoparticles and fix  
their mutual separation. Standard SERS-active surfaces are well-known
to be difficult to use in a controlled way, and encapsulated nanostructures
can provide, therefore, a novel technique for the preparation
of well-defined SERS-active substrates. Moreover,
the capability of these systems to enhance  
optical fields can be used not only for
spectroscopic purposes: if we shine a laser beam of appropriate frequency
into a solution of Ag-filled carbon nanotubes, van der Waals attraction
between the tubes can be greatly photo-enhanced and binding into 3D
structures of Ag-filled carbon nanotubes can be facilitated.

One of us (J.M.P.) acknowledges financial support by the Basque Hezkuntza, 
Unibertsitate eta Ikerketa Saila and the Spanish Ministerio de Educaci\'on y 
Cultura.

{\Large {\bf Figure Captions}} \newline

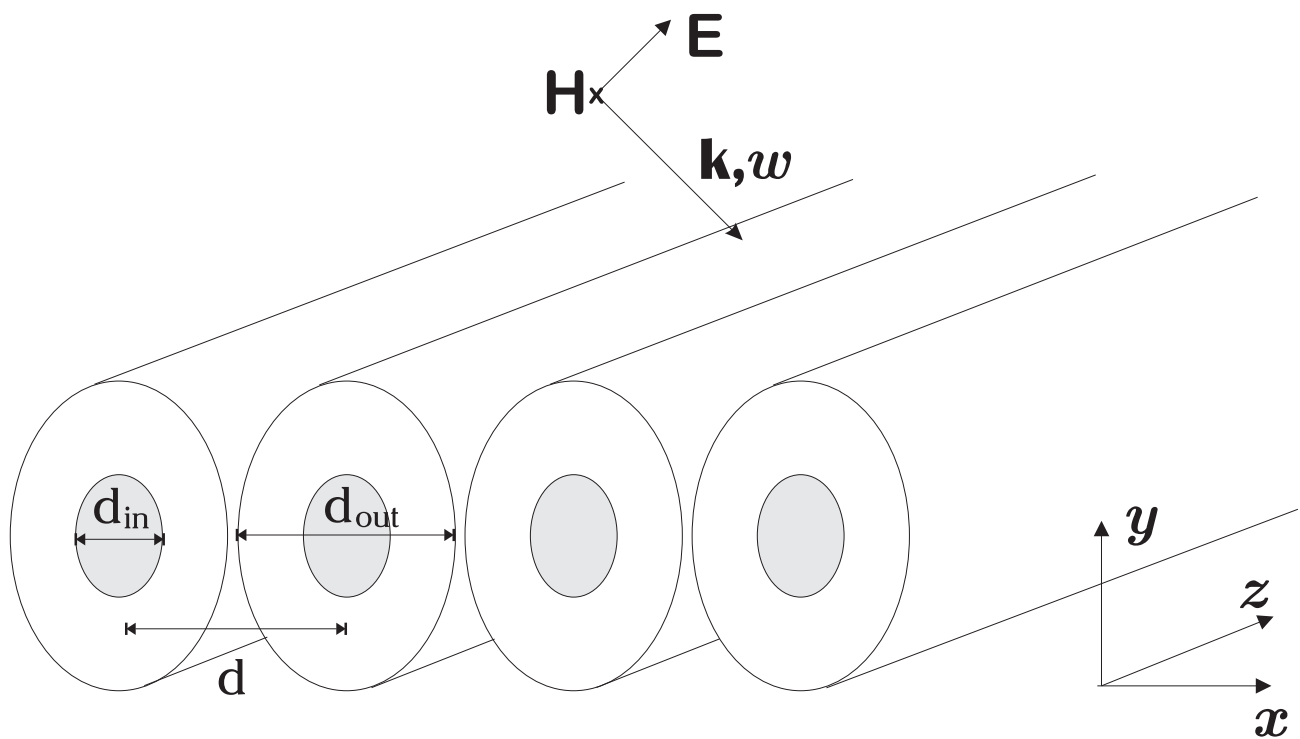
{\bf Figure 1}. Our model of an array of Ag-filled carbon nanotubes: the
cylinders of outer diameter $d_{out}=10$ nm are infinitely long in the $z$%
-direction and are distributed periodically in the $x$-direction with a
separation of $d=10.3$ nm. The inner core filled with silver will be varied
between $3$ and $9.5$ nm in our calculations. We study the electromagnetic
interaction of this structure with a normally incident {\it p-polarized} plane wave of
momentum ${\bf k}$ and energy $\omega$. \newline

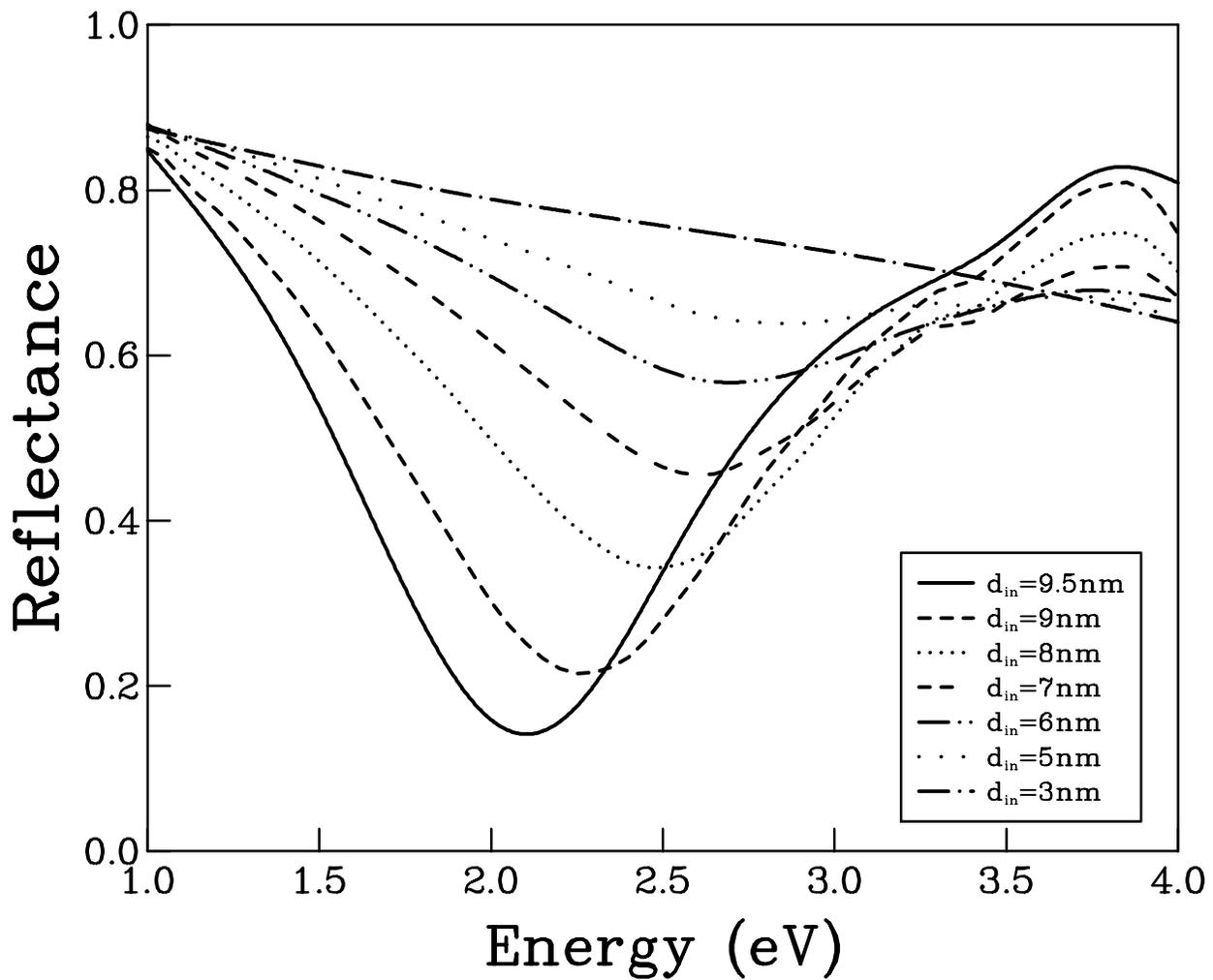
{\bf Figure 2}. Calculated optical reflectance of the structure shown in
Fig. 1 for a normally incident photon of frequency ranging from $1$ to $4$
eV, and different values of the inner core filled with silver,
$d_{in}$%
. \newline

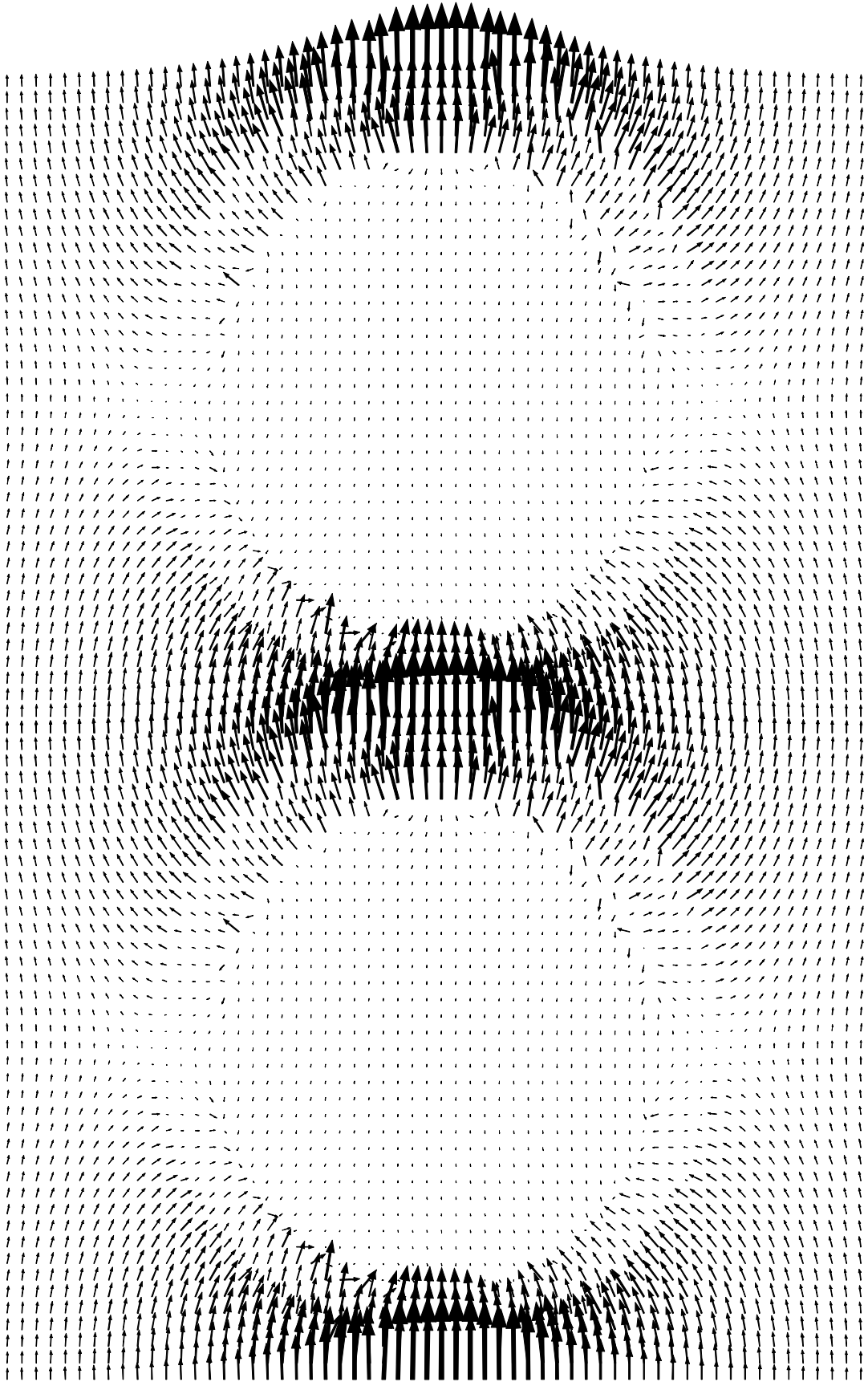
{\bf Figure 3}. A detailed picture of the total {\bf E} field generated by a
normally incident plane wave impinging on the array of Ag-filled carbon
nanotubes of Fig. 1 with $d_{in}=8$ nm. The {\bf E} field is evaluated at $%
\omega=2.5$ eV, the frequency at which the optical reflectance is minimum
for this case. The total {\bf E} field is shown for two unit cells of the array. \newline

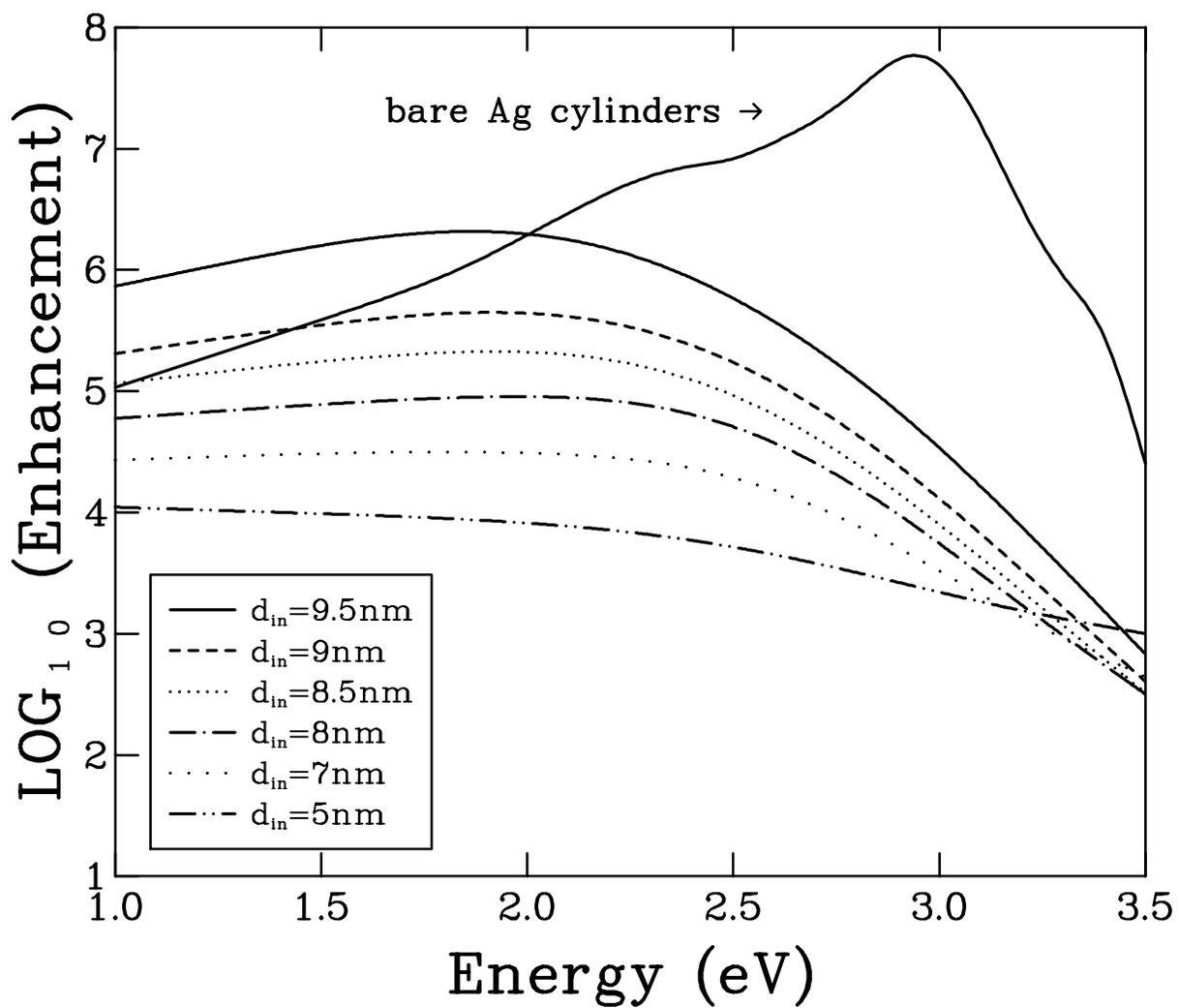
{\bf Figure 4}. The local enhancement of the Raman signal evaluated at the
region between the nanotubes, for different values of $d_{in}$ and as a function
of the energy of the incoming photon. The same 
quantity, as evaluated for bare Ag cylinders of diameter $9.5$ nm and separated 
by $10.3$ nm, is also shown.

\end{document}